\documentclass[aps,prl,twocolumn,showpacs,superscriptaddress,groupedaddress,nofootinbib]{revtex4-1}
\usepackage{amsmath,amsfonts,amssymb,array}
\usepackage{bm,bbm,bbm,bbold}
\usepackage{enumerate}
\usepackage{dsfont}
\usepackage{float}
\usepackage{graphicx}
\usepackage{longtable}
\usepackage{multirow}
\usepackage{slashed,subfigure}
\usepackage{xcolor}
\usepackage[colorlinks = true,linkcolor = blue]{hyperref} 
\definecolor{darkgreen}{rgb}{0.0, 0.5, 0.0}
\hypersetup{bookmarks=true,unicode=true,pdftoolbar=true,pdfmenubar=true,
  pdffitwindow=false,pdfstartview={FitH},
  pdftitle={heavyNThreshold},pdfauthor={Ruiz,Spannowsky,Waite},
  pdfsubject={Subject},pdfcreator={Creator},pdfproducer={Producer},
  pdfkeywords={Heavy Neutrinos}{Gluon Fusion}{Resummation},
  pdfnewwindow=true,colorlinks=true,
  linkcolor=blue,citecolor=purple,filecolor=magenta,urlcolor=cyan}

\begin{document}
\leftline{}
\rightline{IPPP/17/97, MCnet-17-23, CP3-17-54}

\title{Searching for processes with invisible particles using a matrix element-based method}

\author{Danilo Enoque Ferreira de Lima}  
\email{dferreir@cern.ch}
\affiliation{Physikalisches Institut, Ruprecht-Karls-Universit\"at Heidelberg, 69120 Heidelberg, Germany}

\author{Olivier Mattelaer}  
\email{olivier.mattelaer@uclouvain.be}
\affiliation{Centre for Cosmology, Particle Physics and Phenomenology (CP3/IRMP)
Universit\'e Catholique de Louvain, 1348 Louvain-la-neuve, Belgium}

\author{Michael Spannowsky}  
\email{michael.spannowsky@durham.ac.uk}
\affiliation{Institute for Particle Physics Phenomenology, Department of Physics, Durham University, Durham DH1 3LE, U.K.}

\date{\today}

\begin{abstract}
We propose a fully flexible method to discriminate between signal and background using a matrix element-based method in the presence of multiple invisible particles. The proposed method performs a mapping of the measured final state onto an observable which improves the separation between signal and background using their matrix elements. To show how performant this generic method is in separating signal from background, we apply it to the prominent partly invisible decay of a Higgs boson into a muon-antimuon pair and two muon-neutrinos via two W bosons. 
\end{abstract}

\maketitle

\section{Introduction}\label{sec:Intro}
The extraction of few interesting signal events from a large number of Standard Model background events is one of the biggest challenges at the Large Hadron Collider (LHC). Depending on the nature and kinematic topology of the signal, different techniques and strategies have been devised to perform this task.

In general, in a first step, observables that are characteristic to the signal have to be constructed. This could entail simple observables, like the transverse momentum of reconstructed objects, e.g. leptons, photons or jets, or the total amount of missing energy, or more sophisticated observables, like jet substructure observables. If the signal is a heavy resonance that decays into electroweak gauge bosons or the top quark, which in turn have a large branching ratio into jets, studying the substructure of jets is a popular way to separate them from large QCD backgrounds \cite{Abdesselam:2010pt,Altheimer:2012mn}. Either kind of observables can then be further processes using increasingly popular multivariate analysis (MVA) techniques, e.g. neural nets or boosted decision trees, to perform an hypothesis test between signal and background.
 
An alternative way of performing such discrimination is to use the measured particles' momenta as direct input to the evaluation of the matrix element of the assumed underlying process, thereby evaluating if the final state was more likely to be produced by the signal or background hypothesis. This approach is called Matrix Element Method (MEM) \cite{Kondo:1988yd,Abazov:2004cs,Artoisenet:2010cn}. As it is based on an analytic/numerical calculation of the process, this method can be directly applied to data and does not require training on Monte-Carlo-generated pseudo-data.

However, while MEM has been used very successfully in a wider range of applications and measurements \cite{ Alwall:2010cq, Andersen:2012kn, Cranmer:2006zs, Alwall:2009sv, Artoisenet:2013vfa, Betancur:2017kqe}, and recent developments extended it to the substructure of jets \cite{Soper:2011cr,Soper:2012pb}, next-to-leading order accuracy \cite{Campbell:2012cz,Campbell:2013hz,Martini:2015fsa,Gritsan:2016hjl,Martini:2017ydu} and even to an arbitrary number of reconstructed final state objects \cite{Soper:2014rya, Englert:2015dlp}, as an all-information approach, it always had its short-comings when multiple invisible objects are present in the final state. 

Here, we propose a fully flexible method to discriminate between signal and background based on the Matrix Element Method in the presence of multiple invisible particles.
The method maps the set of measured final state into a manifold parametrised by the minimal degrees of freedom for a given process. Such new manifold, which we will refer to as a ``minimal hypersurface'', contains the set of all final states that can be produced for a minimal set of degrees of
freedom. Such parametrisation can then be sampled in a unique way to produce
the set of all final states compatible with the observed event.
With such reparametrisation, one can then maximise the matrix element separately for signal and background.
On the one hand, this allows to make an educated guess of the 4-momenta of the invisible particles in the process, and on the other hand it allows to construct a variable $\chi$ as the ratio of the matrix elements that can be used to separate signal from backgrounds.
We would like to emphasize that while this method is based on matrix elements, it is not identical to the so-called Matrix Element Method~\cite{Kondo:1988yd,Abazov:2004cs,Artoisenet:2010cn},
as it makes approximations in order to calculate the final discriminator faster.

Final states with multiple missing energy particles became an increasingly important signature in searches for a plethora of new physics scenarios at the LHC, e.g. searches for dark matter \cite{Aaboud:2017phn,Sirunyan:2017onm}, R-parity conserving supersymmetry \cite{Sirunyan:2017hvp, Aaboud:2017hrg}, large extra dimensions \cite{Chatrchyan:2012tea}, or even anomalous couplings of the Higgs boson \cite{Aaboud:2017bja, Khachatryan:2016whc}. Thus, a flexible method not relying on Monte-Carlo-generated pseudo-data can be readily applied to ongoing searches and measurements at the LHC's multipurpose experiments and increase their discovery potential.

We emphasize that there are several flavours of the matrix element method, such as the ones described in~\cite{Campbell:2012cz},~\cite{Martini:2015fsa},~\cite{Gritsan:2016hjl} or in~\cite{Soper:2014rya}. 
The method we propose is based on the matrix element method, but it is not identical to it. The method proposed here aims at making the procedure easier and faster, while maintaining the ability to separate between signal and background, in the case of missing energy particles.

\section{Description of the method}\label{sec:Method}
\label{sec:desc}

The matrix element method assigns probabilities to signal and background for each event of 
a sample. The most attractive feature of this method is that  it makes
maximal use of both the experimental information and the theoretical model. It associates a weight
to each event based on the value of the matrix element ({\it i.e.}, the scattering amplitude) for that specific final state configuration for each of  the hypotheses.
The weight associated with  an experimental event $x$, given a set of hypotheses $\alpha$, is
\begin{equation}
\label{mem_def}
P_\alpha(x)=\frac{1}{\sigma_{ \alpha}} \int d \Phi(y) |M_{ \alpha}|^2 ( y)  W(x,y)\,,
\end{equation}
where $|M_\alpha|^2( y)$ is  the squared leading-order matrix element, $d \Phi(y)$ is  the phase-space measure,  (including the parton distribution functions) and $\
W(x, y)$ is the transfer function which describes  the evolution of the final state parton-level configuration in  $y$ into a reconstructed event $x$ in the detector. 
It is defined by the conditional probability to observe an experimental event $x$ when the truth parton event is $y$. This function summarises a lot of different physics effect including parton-shower, hadronization, detector resolution and so on~\cite{WBosonHelicityD0}.

The normalization by $\sigma_{ \alpha}$  in Eq.~(\ref{mem_def}) (dubbed the visible cross-section) ensures that $P_\alpha(x)$ is a probability density, $ \int P_\alpha(x) dx=1$, if the transfer function is normalized to one.
As is evident from the definition in Eq.~(\ref{mem_def}), the calculation of each weight involves a non trivial multi-dimensional integration of complicated functions over the phase space. 
Even if the  problem of computing the weights for arbitrary models and processes can be automated, e.g. as implemented in {\sc MadWeight}~\cite{Artoisenet:2010cn}, such calculations remain extremely CPU intensive and are subject to numerical inaccuracies.
We instead propose to replace the convolution of the matrix element with the transfer function by a maximisation procedure over the phase-space volume~\footnote{
Note that in the equation below, the factor $ |M_{ \alpha}|^2W(x,y)$ is not dimensionless, however one can easily renormalise such
quantity by a proportionality constant that cancels out the dimensionful component of the matrix element squared. Such normalisation factor
would cancel out for signal and background events with the same number of unmeasured final state particles.
}
$\Phi$, 
\begin{equation}
\label{w_def}
w_\alpha(x) = \max_{y \in \Phi}\left( |M_{ \alpha}|^2 (y)  W(x,y)\right).
\end{equation}
In order to use efficiently the maximization algorithms over a highly dimensional space, it is important to parametrize the phase-space in an optimal way. In particular, the invariant mass of every propagator that can be on-shell needs to be used as a degree of freedom of the phase-space, as well as all the angles of visible particles (due to the high detector resolution on those quantities). Such parametrization allows to reduce the variance of the function by smoothing the peak and it helps to find its maximum more efficiently. We rely on {\sc MadWeight} to find such a parametrization, which provides a large set of changes of variables that can be combined to reach the optimal parametrization of the phase-space. 

For both Eq. \ref{w_def} and \ref{mem_def}, the amount of CPU time needed for each event will be related to the presence/absence of local maximum in the function which are typically created by some tension between the partons of the phase-space favoured by matrix-element and the one favoured by the transfer-function. Due to this origin, we do not expect a huge number of false maxima and it is quite simple to identify all of them and find the global maximum.
This problem is much more complex in the case of the full phase-space integration where the contribution of each of those phase-space region needs to be correctly evaluated to obtain a good estimator of the weight. Obviously this also means that using only the maximum reduces the information encoded in our final weight and will reduce the sensitivity of the method (as it should  be evident from the Neyman-Pearson Lemma).

After finding the most likely final state configuration, given a limited amount of information~\footnote{That is, only the visible final state, which can be plagued by experimental uncertainties.}, we construct an observable $\chi$, which classifies each event on whether it appears more signal- or background-like:
\begin{equation}
\chi =  \frac{w_S}{\sum_i w_{B_i}}.
\end{equation}
A selection requirement can be applied on $\chi$ to reject background,
improving the analysis sensitivity.
The significant gain in speed and high performance of the classifier, allows one to extend it to complex final states with many objects. The method can be integrated straightforwardly into the {\sc EventDeconstruction} approach \cite{Soper:2014rya}, thereby extending  {\sc EventDeconstruction}, which was already designed to handle an arbitrary number of visible final state objects, to final states with invisible particles. 

Thus, even if this letter focuses on a single example, the method is entirely generic and can be applied to a large class of analyses. We will release a generic code \cite{futurearticle}, which allows to apply the above method efficiently for any process and set of transfer functions, hence providing the same flexibility as {\sc MadWeight}.

We note additionally that the introduction of transfer functions are necessary
when the measured objects have a much worse momentum resolution than required
to map out the fast change of the matrix element. More precisely, for example, in the process $pp \to HZ$ with subsequent decay $H\rightarrow b\bar{b}$,
the matrix element is proportional to the Breit-Wigner propagator $\frac{1}{(p_{b,1} + p_{b,2})^2 - m_H^2 + i m \Gamma_H}$, and is thus maximised when $(p_{b,1} + p_{b,2})^2 = m_H^2$.
However, because the width of the Higgs boson is only $\sim 4$ MeV and $\sqrt{(p_{b,1} + p_{b,2})^2}$
can experimentally only be reconstructed with a precision of $\mathcal{O}(\mathrm{GeV})$, measurement uncertainties dominate the value of the matrix element. Thus, one introduces transfer functions over which one integrates to ensure that the maximum contribution from the matrix element (i.e. when $(p_{b,1} + p_{b,2})^2 = m_H^2$)
is included in the classification between signal and background. Thus, transfer functions are particularly
useful in the decay of resonances, when the matrix element changes quickly.
However, when the measurement uncertainties on the momenta of particles are small compared
to the change of the matrix element over the change of the momentum they can be neglected.

This motivates the maximisation procedure we propose here. Invisible particles arise in the Standard Model,
and in most envisioned expansions, in decays of electroweak (or heavier) resonances.
Thus, we expect the matrix element to be maximised when $m_W^2 = (p_{\nu} + p_{\ell})^2$ and when
$m_H^2 = (p_{W,1} + p_{W,2})^2$ (in the signal). When the matrix element is peaked in such phase space
regions (which is the case for most processes including invisible particles) our approach is a good approximation to integrating over all degrees of freedom for all possible momenta as the final weight is
dominated by the maximum of the matrix element.

\section{Partly invisible Higgs boson reconstruction}\label{sec:HiggsRec}

To show how performant the method is in separating signal from background, we apply it to the prominent partly invisible decay of a Higgs boson into a muon-antimuon pair and two muon-neutrinos via two W bosons \cite{Dittmar:1996ss, Khachatryan:2016vnn, ATLAS:2014aga}. 
The $pp \to H \to W^+ W^- \to \mu^+ \nu_\mu \mu^- \bar{\nu}_\mu$ signal and dominant background \cite{Chatrchyan:2013iaa}, $pp \to W^+ W^- \to \mu^+ \nu_\mu \mu^- \bar{\nu}_\mu$, have been simulated using MadGraph5\_aMC@NLO 2.5.2~\cite{Alwall:2014hca,Hirschi:2015iia} and showered with Pythia 8.226~\cite{Sjostrand:2007gs}, thereby allowing for hadronisation effects and additional initial state radiation.
We assume an integrated luminosity of 30 fb$^{-1}$ and simulate proton-proton collisions at $\sqrt{s} = 13$ TeV.

Before we apply the matrix-element-method, we select candidate events with the following event selection cuts, which render all but one irreducible background process insignificant. We select muons with a minimum transverse momentum requirement of $p_{T,\mu} > 10$ GeV and a requirement that
the absolute value of the pseudo-rapidity is $|\eta_\mu | < 2.5$, to ensure that the muons are within the range of the detector's tracker system. The experimental resolution on the momenta of the muons are precise enough for us to assume their experimental uncertainty to be negligible. Thus, we define $W(x,y)$ of Eq.~\ref{w_def} as $W(x,y)=\delta^4(p_{\mu^+}^{exp}-p_{\mu^+}^{MC})\delta^4(p_{\mu^-}^{exp}-p_{\mu^-}^{MC})$. To reduce the muon mis-identification rate, the sum $I$ of charged
particles within $\Delta R(\mu, \textrm{particle}) < \min(0.3, 10\textrm{ GeV}/p_T^{\mu})$ has been required to satisfy $I/p_T^{\mu} < 0.06$.

\begin{table}[ht]
\begin{tabular}{c|c|c|c|c}
                                     & Signal            & Background       & $s/b$          & $s/\sqrt{b}$        \\
\hline
Basic event selection cuts           & 327               & 11451            & 0.029          & 3.058               \\
\hline
\multicolumn{5}{c}{Assuming perfect $E_T^{\rm miss}$ reconstruction} \\
\hline
Veto $w_{S,B}(S,B) = 0$            & 299               & 3724             & 0.080          & 4.912               \\
$\chi > 10$                     & 262               & 2200             & 0.119          & 5.592               \\
$\chi > 31$                   & 118               & 808              & 0.146          & 4.157               \\
\hline
\multicolumn{5}{c}{Assuming a 10\% resolution effect in $E_T^{\rm miss}$} \\
\hline
Veto $w_{S,B}(S,B) = 0$              & 294               & 3742             & 0.079          & 4.806               \\
$\chi > 10$                     & 256               & 2204             & 0.116          & 5.455               \\
$\chi > 31$                   & 114               & 811              & 0.141          & 4.016               \\
\end{tabular}
\caption{Signal-to-background ratio and signal-to-square-root-of-background ratio after basic selection and the proposed method.}
\label{tab:sb}
\end{table}

With these basic selection cuts, for $\sqrt{s} = 13$ TeV and an integrated luminosity of $L=30$ fb$^{-1}$, and no detector simulation, we obtain a signal-to-background ratio of $S/B \simeq 0.03$ and a statistical sensitivity of $S/\sqrt{B} \simeq 3.06$, as shown in Table~\ref{tab:sb}.

\begin{figure}[ht]
\centering
\includegraphics[width=0.8\linewidth]{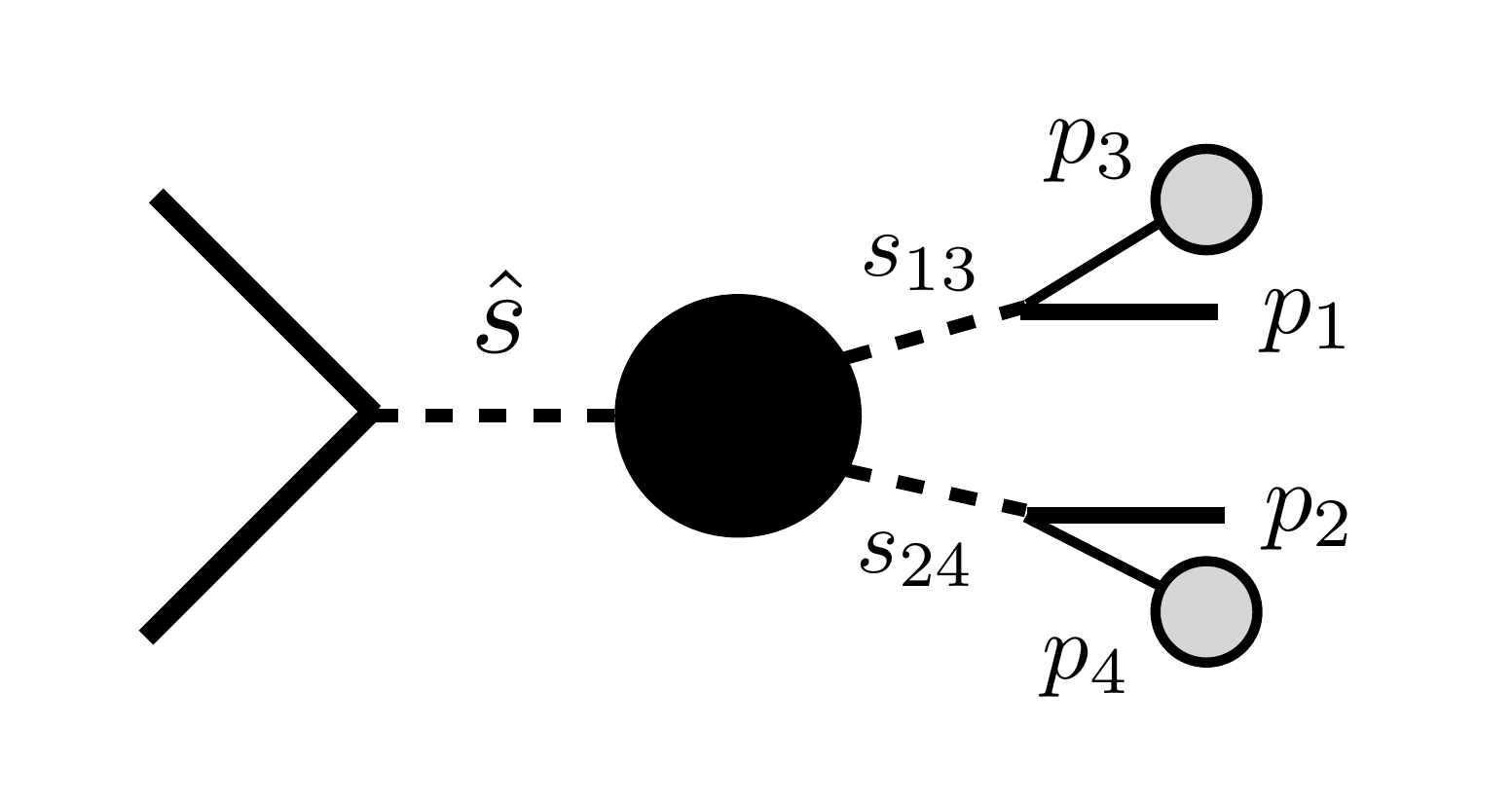}
\caption{Kinematic for signal and background processes.}
\label{fig:kin}
\end{figure}

For the signal pseudo-data generated, we can now use the method discussed
to evaluate the weight for a signal event to look like signal $w_S(S)$ or to look like background $w_B(S)$. The impact of initial state radiation is dampened by implementing a boost back technique of the reconstructed
momenta of the lepton as suggested in \cite{Alwall:2010cq}. The four free parameters defining the phase space $\Phi$ over which we maximise the matrix elements are $\hat{s}$, $s_{13}$, $s_{24}$, as shown in Fig.~\ref{fig:kin}, and the rapidity of the full system $y_\mathrm{all}$, where all includes the two muons and the missing transverse energy. In the example at hand, we can impose further boundary conditions, i.e. $\sqrt{\hat{s}} \simeq m_h$, $\sqrt{s_{13}} \simeq m_W$ and  $\sqrt{s_{24}} < m_W$ for the signal and  $2 m_W < \sqrt{\hat{s}} < 3 m_W$, $\sqrt{s_{13}} \simeq m_W$ and  $\sqrt{s_{24}} \simeq m_W$ for the background\footnote{We tested larger windows for $\sqrt{\hat{s}}$ but did not find them to change the background weights significantly. The asymmetric phase-space cuts on $\sqrt{s_{13}}$ and $\sqrt{s_{24}}$ are flipped half of the time when maximizing over the phase space. }.

Despite limiting the four-dimensional parameter space, the matrix-element weighted hypersurface is complicated enough to give rise to multiple minima or to fail to give a physical solution for the matrix element entirely.
As we probe the parameter space spanned by the unmeasured degrees of freedom, subject to restrictions mentioned previously,
it may happen that no probed parameter space satisfies such restrictions.
This is a consequence of using Leading-Order matrix-element and simplified transfer function.
A detailed study can relate (most of) those events without any physical solution to the presence of radiation
not correctly handled by the transfer function.
Thus, to find the global maximimum we rerun the maximisation procedure with randomly modified initial conditions $n_r = 500$ times for signal and background each\footnote{We have varied $n_r$ between 0 and 500 and find for $n_r > 150$ the change of $w_S$ and $w_B$ to be insignificant.}.

\begin{figure}[ht]
\centering
\subfigure[][]{\includegraphics[width=0.75\linewidth]{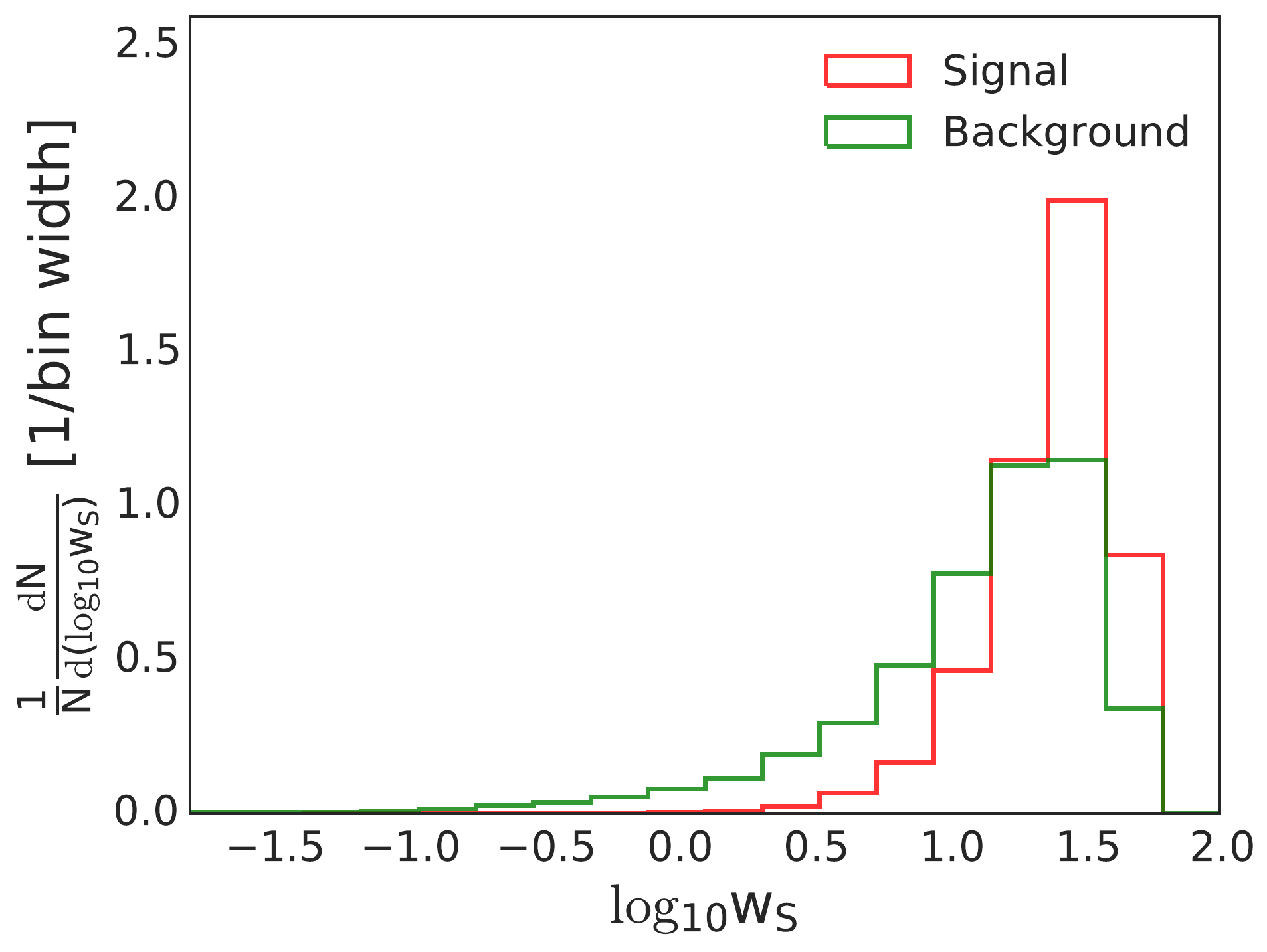}\label{fig:wsig}}\\
\subfigure[][]{\includegraphics[width=0.75\linewidth]{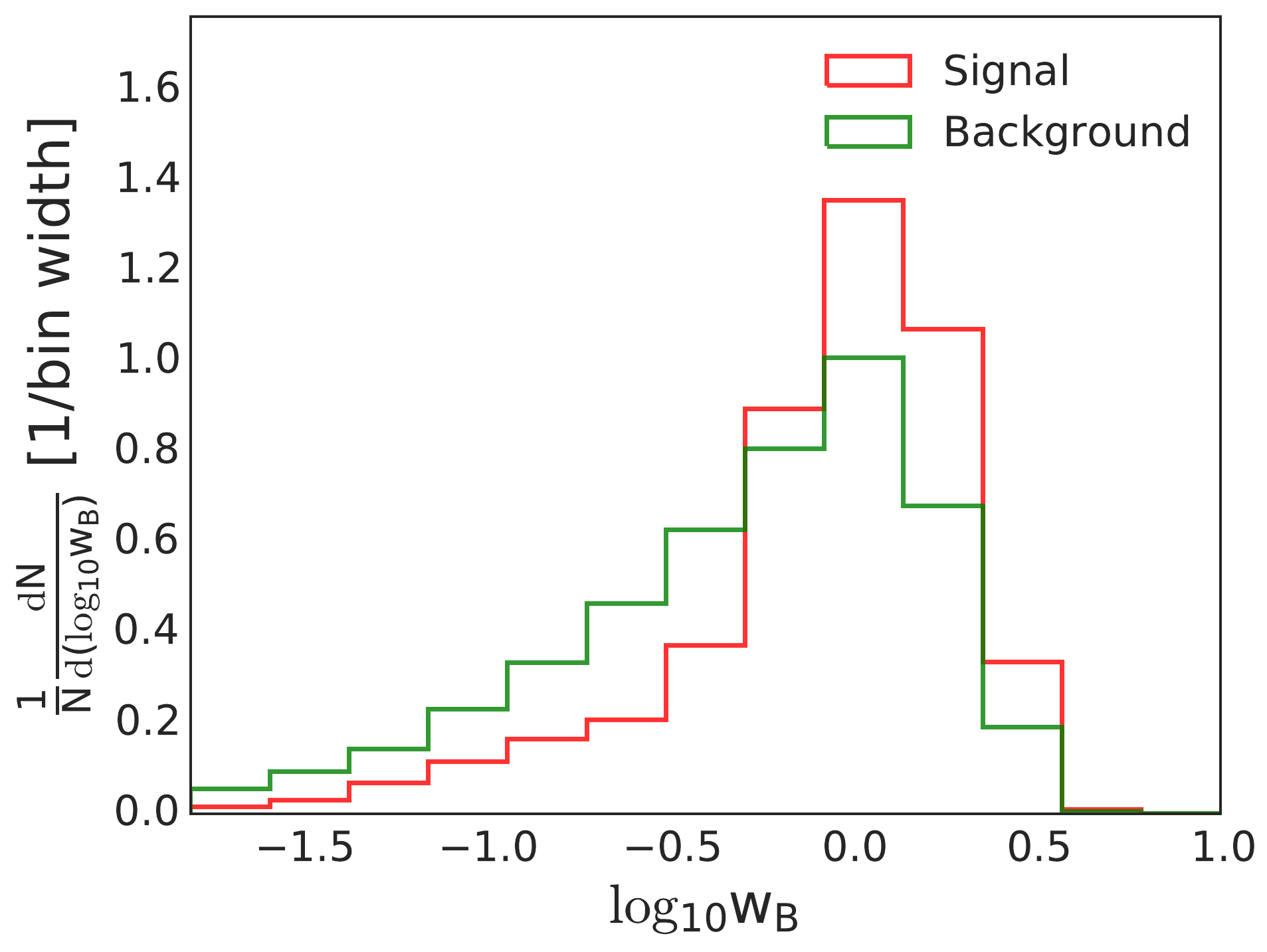}\label{fig:wbkg}}\\
\subfigure[][]{\includegraphics[width=0.75\linewidth]{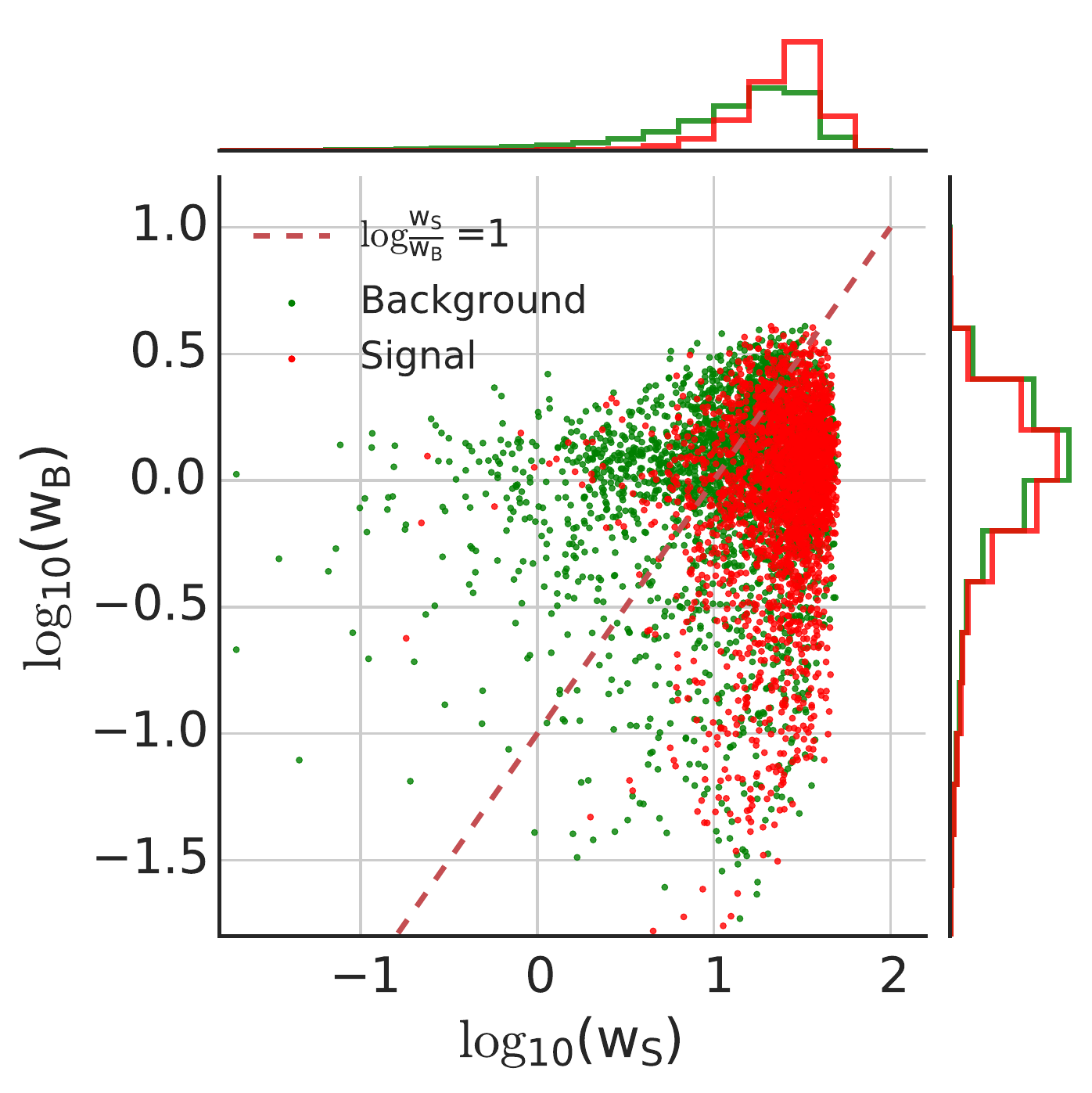}\label{fig:sig_vs_bkg}}
\caption{Signal ($w_S$) and background ($w_B$) weight distributions for signal (red) and background (green) samples respectively.}
\label{fig:wsb_split}
\end{figure}

Thus, we can calculate the weight for the signal and background hypotheses $w_S$ and $w_B$, respectively for signal and background events. We show all four distributions in Figs.~\ref{fig:wsig} and \ref{fig:wbkg}.
Event kinematics which do not result in a physical configuration for the signal or background hypothesis give either $w_S=0$ or $w_B=0$. We do not show such events in Figs.~\ref{fig:wsig} and \ref{fig:wbkg}, but their fraction can be inferred from Table~\ref{tab:sb}.
A fairly large number of background events fail to pass the kinematic requirements to look like signal, i.e. resulting in $w_{S}(B) = 0$. This behaviour is beneficial for the significance of the analysis, as such background events have zero probability to mimic the signal.

\begin{figure}[ht]
\centering
\includegraphics[width=0.8\linewidth]{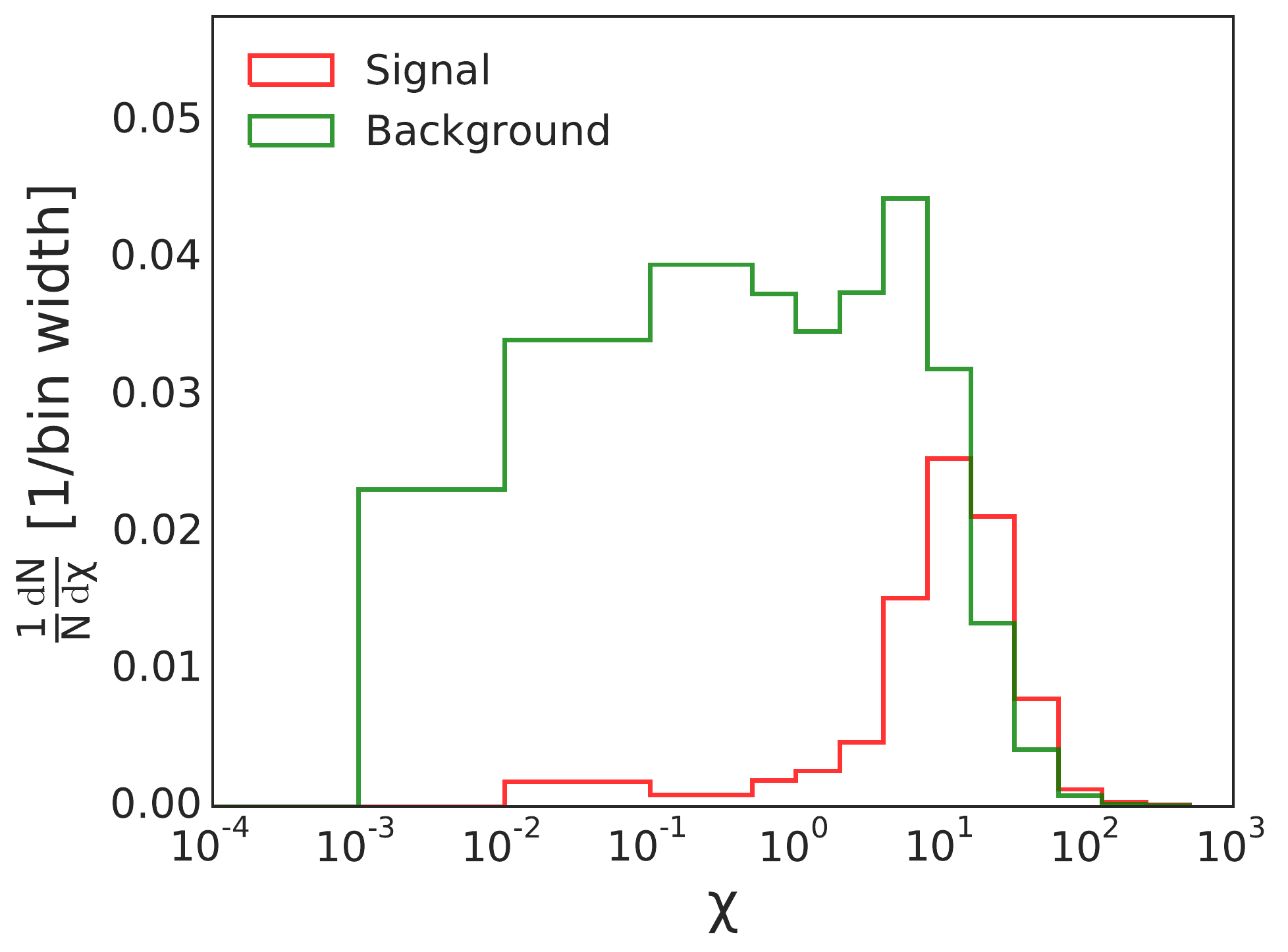}
\caption{Distribution of $\chi$ for signal (red) and background (green).}
\label{fig:wsb}
\end{figure}

\begin{figure}[ht]
\centering
\includegraphics[width=0.8\linewidth]{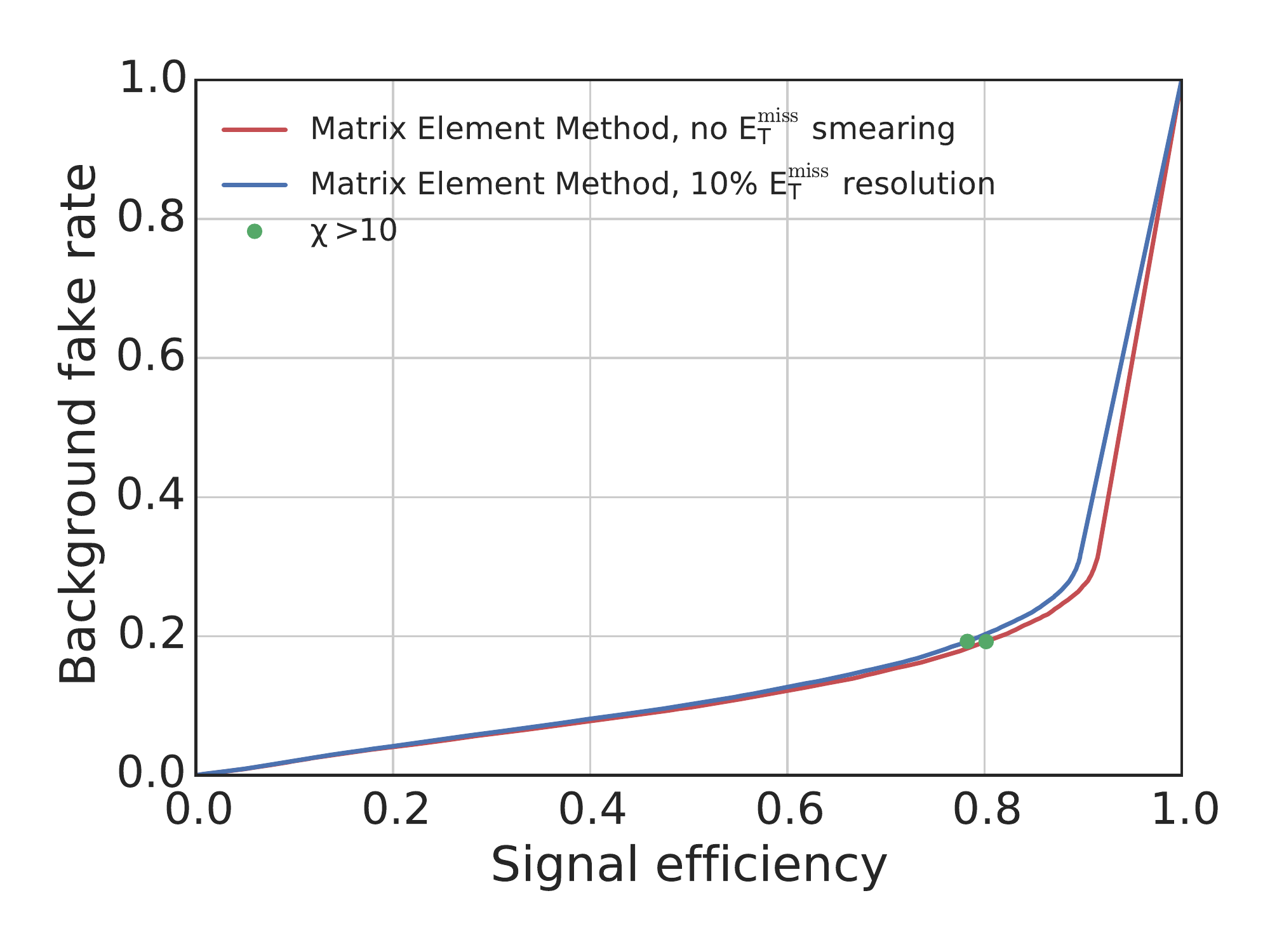}
\caption{Background mis-identification rate versus signal efficiency for the proposed method, with (blue) and without (red) smearing of the missing transverse energy.}
\label{fig:roc}
\end{figure}

After vetoing all events where $w_{S,B}(S,B) = 0$ we find $S/B = 0.08$, $S/\sqrt{B} \simeq 4.91$ and show
the distribution of weights in Fig.~\ref{fig:sig_vs_bkg}. While a comparison of the absolute weights for
the signal and background hypothesis does not allow for a strong separation on an event-by-event basis
(see the distributions on the horizontal and vertical axes of Fig.~\ref{fig:sig_vs_bkg}), taking the ratio 
\begin{equation}
\chi = \frac{w_S}{w_B} 
\end{equation}
for each event results in a strong discrimination between signal and background, see Fig.~\ref{fig:wsb}.
For example, by requiring $\chi \geq 10$ we reject $81 \%$ of background while still accepting $80 \%$ of signal, resulting in $S/B \simeq 0.12$ and $S/\sqrt{B} \simeq 5.59$.
In order to estimate the impact of a selection requirement in the observable $\log(\chi)$, we scan all possible cuts in this variable and estimate
the fraction of accepted background and the signal efficiency in each scenario. The resulting curve, showing the fraction of the accepted background
events in the $Y$ axis and the fraction of accepted signal events in the $X$ axis after a cut in $\log(\chi)$ is referred to as a ``Receiver Operating Curve'' (ROC) and it is shown in Fig.~\ref{fig:roc}.

While the momenta of the charged leptons can be measured very precisely, the total amount of missing transverse energy instead is subject to experimental uncertainties. Such uncertainties can potentially affect the ROC curve and overall significance negatively. To estimate the impact of this uncertainty on our method we include a 10\% resolution effect by smearing the missing transverse energy with a gaussian distribution.

Both in the ROC curve and Table~\ref{tab:sb}, we show the effect of the resolution effect on the missing transverse energy
for this method. We find however, that such effect reduces $s/\sqrt{b}$ only slightly from 5.59 to 5.46.

Instead of a cut and count procedure, one can use the full shape of the $\chi$ distribution of Fig.~\ref{fig:wsb} to set a CLs \cite{Feldman:1997qc, Junk:1999kv} limits on the Higgs-W coupling.
Note that, in this procedure, the $\chi$ variable is never used for the statistical interpretation directly, since it is biased.
Instead, we histogram the variable $\chi$ for signal (under several $g_{H,WW}$ conditions), add it to the histogram of the background
and use the bin contents (under a chosen binning) to define a likelihood function given by the product of Poisson functions with means
at the resulting histogram bin contents. Such procedure is also done for the background-only hypothesis to define a background-only
likelihood. One can then define a likelihood-ratio function, as required for the CLs method.
Including the 10\% resolution effect on the missing transverse energy reconstruction, one can set a 95\% CL limit on the Higgs-W-boson coupling at $g_{H,WW} \in [0.65, 1.25] \times g_{H,WW,SM}$. While a direct comparison is difficult due to the different collider energies, the limit we obtain is already better than the one from the full combined 7 and 8 TeV data set for the gluon-fusion Higgs production process with subsequent decay into W bosons \cite{Aad:2015gba}.

\section{Summary and Conclusion}\label{sec:conclusions}

We have proposed a matrix-element method, designed to perform a hypothesis-test in the presence of multiple invisible particles in the final state. Without integrating over the phase space the most-likely kinematic configuration is calculated separately for signal and background. We make full use of the information available on the particle involved in the process. 

We applied this method to separate the process $pp \to H \to WW^* \to \mu^+ \mu^- \nu_\mu \bar{\nu}_\mu$ from the irreducible background $pp \to WW \to \mu^+ \mu^- \nu_\mu \bar{\nu}_\mu$. Using only objects that are experimentally well under control, i.e. the momenta of the muons, from which we calculate the missing transverse energy, we are able to set a strong limit on the Higgs-coupling to W bosons, assuming an integrated luminosity of $30~\mathrm{fb}^{-1}$ at $\sqrt{s}=13$ TeV.

Other methods to reconstruct partly invisible final states have been devised before, e.g. $m_{T_2}$ \cite{Lester:1999tx, Betancur:2017kqe} or boosted kinematics \cite{Englert:2012wf}. However, this matrix element method is not relying on a specific kinematic structure for the decays, e.g. the presence of particles with the same mass, or the number of invisible final state particles. We will release a generic Monte-Carlo implementation of this method in a future publication \cite{futurearticle}, thereby showing the flexibility and applicability to a wide range of beyond the Standard Model scenarios.

\noindent 
\acknowledgements{
\small{
\textit{
Acknowledgements: \\
OM would like to thank Pierre Artoisenet for fruitful discussions and  the CERN TH division for its hospitality. OM was partly supported by the Belgian Pole d'attraction Inter-Universitaire (PAI P7/37) and by the European Union's Horizon 2020 research and innovation programme as part of the Marie Sklodowska-Curie Innovative Training Network MCnetITN3 (grant agreement no. 722104).
DEFL has been partly supported by the Alexander von Humboldt Foundation.
}}}

\bibliography{references}


\end{document}